\documentclass[article]{JHEP3}
\usepackage{amsmath,amssymb}
\usepackage[dvips]{graphics}
\usepackage{epsfig}
\usepackage{psfrag}

\newcommand{\be}{\begin{equation}}
\newcommand{\ee}{\end{equation}}
\def\ii{{\rm i}}
\newcommand{\beq}{\begin{eqnarray}}
\newcommand{\eeq}{\end{eqnarray}}
\newcommand{\beean}{\begin{eqnarray*}}
\newcommand{\eean}{\end{eqnarray*}}

\relax

\title{Superradiant instabilities of rotating black branes and strings}

\author{Vitor Cardoso$^{\dag,\ddag}$
and Shijun Yoshida $^{\dag\dag}$\\
$^{\dag}$McDonnell Center for the Space Sciences,
Department of Physics, Washington University, St. Louis, Missouri 63130, USA\\\\
$^{\ddag}$Centro de F\'{\i}sica Computacional, Universidade de
Coimbra,
P--3004--516 Coimbra, Portugal\\\\
$^{\dag\dag}$Science and Engineering,
Waseda University, Okubo, Shinjuku, Tokyo 169-8555, Japan\\\\
\email{vcardoso@wugrav.wustl.edu}, \quad\email{shijun@waseda.jp}}

\abstract{Black branes and strings are generally unstable against a
certain sector of gravitational perturbations. This is known as the
Gregory-Laflamme instability. It has been recently argued
\cite{marolf,cl} that there exists another general instability
affecting many rotating extended black objects. This instability is
in a sense universal, in that it is triggered by any massless field,
and not just gravitational perturbations. Here we investigate this
novel mechanism in detail. For this instability to work, two
ingredients are necessary: (i) an ergo-region, which gives rise to
superradiant amplification of waves, and (ii) ``bound'' states in
the effective potential governing the evolution of the particular
mode under study. We show that the black brane ${\rm Kerr}_4 \times
R ^{\,{\rm p}}$ is unstable against this mechanism, and we present
numerical results for instability timescales for this case. On the
other hand, and quite surprisingly, black branes of the form ${\rm
Kerr}_{\rm d}\times R ^{\,{\rm p}}$ are all stable against this
mechanism for ${\rm d}>4$. This is quite an unexpected result, and
it stems from the fact that there are no stable circular orbits in
higher dimensional black hole spacetimes, or in a wave picture, that
there are no bound states in the effective potential. We also show
that it is quite easy to simulate this instability in the laboratory
with acoustic black branes. }

\begin{document}

\section{Introduction}
We are all painfully aware that exact, closed form solutions to
physical problems are an exception. The rule is that one must resort
to all kinds of tricks and assumptions in order to get a grip on the
equations: this is why, despite their little insight into the
physics, numerical solutions are proliferating as problems become
more and more complex. Exact solutions are therefore most welcome.
Exact solutions to Einstein equations are extremely useful,
especially if they describe simple yet physically attainable
systems. Indeed, take for example the famous Schwarzschild metric:
with this exact solution at hand, describing the geometry outside a
spherically symmetric distribution of matter, one was able to
compute the deflection of light as it passes near the Sun (and to
match the theoretical prediction against the observational data),
thereby giving strong support to Einstein's theory. We now know that
the outside geometry of many astrophysical objects is well described
by the Schwarzschild metric, and we can start studying them by
investigating the properties of this metric.

One of the most important things that one should study first is the
classical stability of a given solution: if a solution is unstable,
then it most certainly will not be found in nature (unless the
instability is secular) and the solution loses most of its power.
What does one mean by stability? In this classical context,
stability means that a given initially bounded perturbation of the
spacetime remains bounded for all times. For example, the
Schwarzschild spacetime is stable against all kinds of
perturbations, massive or massless \cite{regge}. Thus, if one
considers small deviations from the Schwarzschild geometry (for
instance throwing a small stone into the black hole, this disturbs
the geometry), stability means that these small perturbations
eventually fade away. They will never disrupt the spacetime. The
term ``perturbations'' means either small (metric) deviations in the
geometry or small-amplitude fields in the geometry, for example
scalar fields or electromagnetic fields (as such, deviations in the
geometry can be looked at as perturbations induced by gravitons).
So, the Schwarzschild geometry is indeed appropriate to study
astrophysical objects.

Here we shall investigate an instability which is
rotation-triggered, and which was hinted at for the first time by
Marolf and Palmer \cite{marolf}. The mechanism has recently been
explained in \cite{cl}, where it was shown that it could be
triggered in many systems. However, a detailed description is still
lacking, and this is the main purpose of the present work. The
physical nature of this instability can be understood with the
following facts: It is known that the Kerr geometry displays
superradiance \cite{zelmisnerunruhbekenstein}. This means that in a
scattering experiment of a wave with frequency $\omega<m\Omega$ ($m$
is again the azimuthal wave quantum number) the scattered wave will
have a larger amplitude than the incident wave, the excess energy
being withdrawn from the object's rotational energy. Now suppose
that one encloses the rotating black hole inside a spherical mirror.
Any initial perturbation will get successively amplified near the
black hole event horizon and reflected back at the mirror, thus
creating an instability. This is the black hole bomb, as devised in
\cite{teupress} and recently improved in \cite{cardosoetal}. This
instability is caused by the mirror, which is an artificial wall,
but one can devise natural mirrors if one considers massive fields
\cite{damour,massivescalar}. Imagine a wavepacket of the massive
field in a distant circular orbit. The gravitational force binds the
field and keeps it from escaping or radiating away to infinity. But
at the event horizon some of the field goes down the black hole, and
if the frequency of the field is in the superradiant region then the
field is amplified. Hence the field is amplified at the event
horizon while being bound away from infinity. Yet another way to
understand this, is to think in terms of wave propagation in an
effective potential. If the effective potential has a well, then
waves get ``trapped'' in the well and amplified by superradiance,
thus triggering an instability. In the case of massive fields on a
(four-dimensional) Kerr background, the effective potential indeed
has a well. Consequently, the massive field grows exponentially and
is unstable. It is the presence of a bound state that simulates the
mirror, and without a bound state we should never get an
instability.

Now, consider for example the following rotating black-brane
geometry \be ds^2=ds^2_{\rm Kerr}+dx^idx_i\,,\label{blackkerr} \ee
where ${\rm Kerr}$ stands for the usual Kerr geometry. It's a known
property, and we will show it explicitly, that the propagation of a
{\it massless} field (scalar, electromagnetic, or gravitational) in
this {\it black-brane} geometry, is {\it equivalent} to the
propagation of a {\it massive} field in the vicinity of the Kerr
{\it black hole}. Thus, the particular black brane (\ref{blackkerr})
is unstable.

The instability argument applies to many rotating extended objects,
and here we shall study some of them, making an extensive analysis
of the ${\rm d}-$dimensional case (\ref{blackkerr}). We show that
for ${\rm d}=4$ they are unstable, and we present detailed results
on the stability. For ${\rm d}>4$ the black branes described by \be
ds^2=ds^2_{{\rm Kerr}_{\rm d}}+dx^idx_i\,,\label{blackkerrD} \ee are
stable, where Kerr$_{\rm d}$ is the Myers-Perry \cite{myersperry}
rotating black hole. This is due to the non-existence of stable
bound orbits for massive particles or, in terms of wave propagation,
there is no well in the effective potential for these systems. To
conclude, we show that one can mimic this instability in the
laboratory using analogue acoustic black branes.
\section{Rotating Kerr-like black branes}
In this section we study in detail the black branes of the form
(\ref{blackkerrD}). In higher dimensions there are several choices
for rotation axes of the Myers-Perry solution (labeled ${\rm
Kerr}_{\rm d}$ in (\ref{blackkerrD})) and there is a multitude of
angular momentum parameters, each referring to a particular rotation
axis \cite{myersperry}. We shall concentrate on the simplest case,
for which there is only one angular momentum parameter, denoted by
$a$.

\subsection{Formalism}

In Boyer-Lindquist-type coordinates the black branes we study in
this section are described by
\begin{eqnarray}
ds^2&=& -{\Delta-a^2\sin^2\theta\over\Sigma}dt^2
-{2a(r^2+a^2-\Delta)\sin^2\theta \over\Sigma}
dtd\varphi +\nonumber\\
&&{}{(r^2+a^2)^2-\Delta a^2 \sin^2\theta\over\Sigma} \sin^2\theta
d\varphi^2 +{\Sigma\over\Delta}dr^2
+{\Sigma}d\theta^2+r^2\cos^2\theta d\Omega_{n}^2+dx^idx_i,
\label{metric}
\end{eqnarray}
where
\be \Sigma=r^2+a^2\cos^2\theta\,,\,\,\, \Delta=r^2+a^2-M r^{1- n},
\ee
and $d\Omega_{n}^2$ denotes the standard metric of the unit
$n$-sphere ($n={\rm d}-4$), the $x^i$ are the coordinates of the
compact dimensions, and $i$ runs from $1$ to $p$. This metric
describes a rotating black brane in an asymptotically flat, vacuum
space-time with mass and angular momentum proportional to $M$ and $M
a$, respectively. Hereafter, $M,a>0$ are assumed.

The event horizon, homeomorphic to $S^{2+ n}$, is located at
$r=r_+$, such that $\Delta|_{r=r_+}=0$.  For $n=0$, an event horizon
exists only for $a<M/2$. When $n=1$, an event horizon exists only
when $a<\sqrt{M}$, and the event horizon shrinks to zero-area in the
extreme limit $a\rightarrow\sqrt{M}$. On the other hand, when $n\ge
2\,,$ $\Delta=0$ has exactly one positive root for arbitrary $a>0$.
This means there is no bound on $a$, and thus there are no extreme
Kerr black branes in higher dimensions.

Consider now the evolution of a massless scalar field $\Psi$ in the
background described by (\ref{metric}). The evolution is governed by
the curved space Klein-Gordon equation \be \frac{\partial}{\partial
x^{\mu}} \left(\sqrt{-g}\,g^{\mu \nu}\frac{\partial}{\partial
x^{\nu}}\Psi \right)=0\,, \label{klein} \ee where $g$ is the
determinant of the metric. We can simplify considerably equation
(\ref{klein}) if we separate the angular variables from the radial
and time variables, as is done in four dimensions \cite{brill}. This
separation was accomplished, for higher dimensions, in \cite{frolov}
for five dimensional Kerr holes and also in \cite{page} for a
general $4+n$-dimensional Kerr hole. Since we are considering only
one angular momentum parameter, the separation is somewhat
simplified, and we can follow \cite{ida}. In the end our results
agree with the results in \cite{frolov,page} with only one angular
momentum parameter in their equations.

We consider the ansatz $\phi=e^{-\ii \omega t+\ii m\varphi+\ii \mu
_i x^i}R(r)S(\theta)Y(\Omega)$, and substitute this form in
(\ref{klein}), where $Y(\Omega)$ are hyperspherical harmonics on the
$n$-sphere, with eigenvalues given by $-j(j+n-1)$
($j=0,1,2,\cdots$). Then we obtain the separated equations \be
{1\over\sin\theta\cos^n\theta}\left({d\over d \theta}
\sin\theta\cos^n\theta{dS\over d \theta}\right)
+\left[a^2(\omega^2-\mu ^2)\cos^2\theta -m^2\csc^2\theta
-j(j+n-1)\sec^2\theta +A\right]S=0, \label{ang} \ee and
\begin{eqnarray}
&&r^{-n}{d\over d r}\left(r^n\Delta{dR\over d r}\right) + \left\{
{\left[\omega(r^2+a^2)-ma\right]^2\over\Delta} \right.
{}\left. -{j(j+n-1)a^2\over r^2} -\lambda-\mu ^2r^2
\right\}R=0, \label{rad}
\end{eqnarray}
where $A$ is a constant of separation, $\lambda:=A-2m\omega
a+\omega^2 a^2$, and $\mu ^2=\sum \mu _{i}^2$. Interestingly, note
the important point that Equations (\ref{ang})-(\ref{rad}) are just
those that describe the evolution of a massive scalar field, with
mass $\mu$, in a ${\rm d}-$dimensional Kerr geometry.

The equations (\ref{ang}) and (\ref{rad}) must be supplemented by
appropriate boundary conditions, which are given by
\begin{equation}
R\sim\left\{
\begin{array}{ll}
(r-r_+)^{-\ii \sigma} & {\rm as}\ r\rightarrow r_+ \,, \\
r^{-(n+2)/2} {\rm e}^{\ii {\sqrt{\omega^2-\mu ^2} r}} & {\rm as}\
r\rightarrow \infty\,,
\end{array}
\right. \label{bound}
\end{equation}
with
\begin{equation}
\sigma \equiv{\left[(r_+^2+a^2)\omega-ma\right]r_+ \over
(n-1)(r_+^2+a^2)+2r_+^2} \,.\label{defb}
\end{equation}
In other words, the waves must be purely ingoing at the horizon and
purely outgoing at the infinity. For assigned values of the
rotational parameter $a$ and of the angular indices $l\,,j\,,m$
there is a discrete (and infinite) set of frequencies called
quasinormal frequencies, QN frequencies or $\omega_{QN}$, satisfying
the wave equation (\ref{rad}) with the boundary conditions just
specified by Eq. (\ref{bound}). Since the time-dependence of the
mode is $e^{-\ii \omega t}$, unstable modes will have frequencies
with a {\it positive} imaginary part, and thus grow exponentially
with time.
\subsection{The physical nature of the instability}
The equations presented in the previous section are amenable to
numerical calculations ({\it infra}), but don't offer much physical
insight into the physics. To gain some more intuition, and
understand {\it why} we expected an instability to be triggered, we
find it convenient to make instead the following change of
variables, following Furuhashi and Nambu \cite{massivescalar}: \be
R=r^{-n/2-1}\Psi \,,\ee in which case the wave equation (\ref{rad})
is transformed into \be \frac{d^2 \Psi}{dz_*^2}-V
\Psi=0\,.\label{wave} \ee Here, we have defined the modified
tortoise coordinate $z_*$ as $\frac{dr}{dz_*}=\frac{\Delta}{r^2}$.
(Note that the usual definition of the tortoise coordinate $r_*$ is
$dr/dr_*=\Delta/(r^2+a^2)$).

 The effective potential $V$ is
equal to \beq  -V &=&\frac{1}{r^2} (-\lambda-a^2\mu
^2-\frac{n}{2}-\frac{n^2}{4}-2am\omega+2a^2\omega ^2)+
\frac{a^2}{r^4}(a^2\omega
^2-2am\omega-\frac{n^2}{2}-jn+m^2-\lambda-j^2+j+2) +\nonumber \\
& &\frac{a^4}{r^6}(2+j-j^2+\frac{n}{2}-nj-\frac{n^2}{4})+\frac{\mu
^2M}{r^{n+1}}+\frac{M}{r^{n+3}}(\lambda-1-\frac{n}{2})+\frac{M
^2}{r^{n+4}}(1+n+\frac{n^2}{4}) +\nonumber
\\ & &
\frac{M a^2}{r^{n+5}}(j^2-j-3+j n+\frac{3n}{2})+\omega ^2 -\mu ^2
\,.\label{effV}\eeq

Upon close inspection, this potential has two key ingredients which
can trigger the instability: superradiant amplification of scattered
waves and the existence of bound states. We shall briefly explore
and describe these features in what follows.

\subsubsection{Superradiant scattering}
In a scattering experiment (with $\omega>\mu$ so that plane waves at
infinity are possible), equation (\ref{wave}) has the following
asymptotic behavior:
\begin{equation}
\Psi _1 \sim\left\{
\begin{array}{ll}
T(r-r_+)^{-\ii \sigma} & {\rm as}\ r\rightarrow r_+ \,, \\
R{\rm e}^{\ii {\sqrt{\omega^2-\mu ^2} r}}+ {\rm e}^{-\ii
{\sqrt{\omega^2-\mu ^2} r}}& {\rm as}\ r\rightarrow \infty\,.
\end{array}
\right. \label{bound2}
\end{equation}
where $\sigma$ was defined in the previous section, equation
(\ref{defb}).

These boundary conditions correspond to an incident wave of unit
amplitude from $+\infty$ giving rise to a reflected wave of
amplitude $R$ going back to $+\infty$ and a transmitted wave of
amplitude $T$ at $-\infty$ (as before, the boundary conditions we
impose do not allow for waves emerging from the horizon). Since the
potential is real, the complex conjugate of the solution $\Psi _1$
satisfying the boundary conditions (\ref{bound2}) will satisfy the
complex-conjugate boundary conditions:
\begin{equation} \Psi _2
\sim\left\{
\begin{array}{ll}
T^*(r-r_+)^{\ii \sigma} & {\rm as}\ r\rightarrow r_+ \,, \\
R^*{\rm e}^{-\ii {\sqrt{\omega^2-\mu ^2} r}}+ {\rm e}^{\ii
{\sqrt{\omega^2-\mu ^2} r}}& {\rm as}\ r\rightarrow \infty\,.
\end{array}
\right. \label{bound3}
\end{equation}
Now, these two solutions are linearly independent, and standard
theory of ODE tells us that their Wronskian is a constant
(independent of $r$). If we evaluate the Wronskian near the horizon,
we get \be W= -2\ii \sigma A|T|^2\,\ee where the constant $A$ is
defined by \be A= \frac{\Delta}{r^2(r-r_+)}\,\,,r \rightarrow
r_+\,,\ee and is positive definite. Evaluating the Wronskian at
infinity we get \be W=2\ii\sqrt{\omega^2-\mu^2}(|R|^2-1)\,.\ee
Equating the two we get \be |R|^2=1-\frac{\sigma A}{\sqrt{\omega
^2-\mu ^2}}|T|^2\,.\ee Now, in general $|R|^2$ is less than unity,
as is to be expected. However, for \be
\omega-\frac{ma}{r_+^2+a^2}<0\,,\ee we have that $\sigma$ is
negative, and therefore in this regime (which will be referred to as
the superradiant regime) $|R|^2>1$. This means that one gets back
more than one threw in. Of course the excess energy comes from the
hole's rotational energy, which therefore decreases. As a final
remark, we notice that we have been assuming without loss of
generality $\omega>0$, and thus superradiance only exists for $m>0$,
which are modes co-rotating with the black hole.
\subsubsection{The potential well}
\begin{figure}
\centerline{\mbox{ \psfig{figure=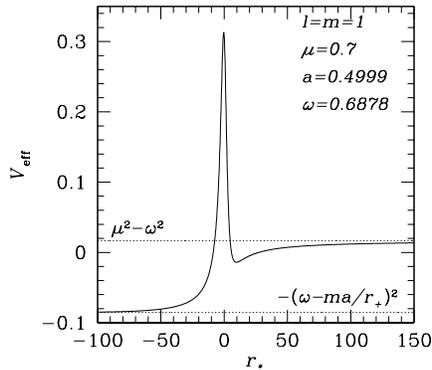,angle=0,width=6cm} }}
\caption{A typical form for the effective potential, here shown for
$l=m=1$ modes. We have set $M=1$, so the rotation parameter $a$
varies between $0$ (Schwarzschild limit) and $1/2$ (extremal limit).
Here we plot the effective potential for the near extreme situation,
$a \sim 0.5$ and for $\mu=0.7$ and $\omega=0.6878$.}
\label{fig:potentialwell}
\end{figure}

The effective potential in four dimensions, ${\rm d}=4$, as given by
(\ref{effV}) is plotted in figure \ref{fig:potentialwell}, for
$a\sim 0.5$ (in units of $M$), $\mu=0.7$ and $\omega=0.6878$.
As can be seen from Figure \ref{fig:potentialwell} the potential
has, in the four-dimensional situation, two extremum between the
event horizon and spatial infinity. The local minimum creates a
``well''-like structure, which will be so important to trigger the
instability. The potential is asymptotic to ($\mu ^2-\omega ^2$) at
spatial infinity. That a well must necessarily arise in four
dimensions can be seen from the asymptotic nature of the potential.
In fact, for $n=0$ and for large $r$, the potential in (\ref{effV})
behaves as \be V  \sim \mu ^2-\omega ^2-\mu ^2M r^{-1}\,\,,\,r
\rightarrow \infty\,\,,\,n=0\,.\ee Thus, the derivative is $V' \sim
\frac{\mu ^2 M}{r^2}$, which is positive. Thus a well necessarily
arises.

The  instability can now be easily explained: the waves get
``trapped'' inside the potential well and amplified by
superradiance. Thus, an exponential growth of any initial
perturbation is inevitable, and therefore an instability arises.
This simple picture is accurate enough that it will allow us to
predict if and when an instability is triggered.
\subsection{The numerical search technique}
The problem we have to solve numerically is a simple boundary value
problem of a second-order ordinary differential equation. However,
this is not easy to solve with the usual techniques of direct
integration, because of difficulties related to the absence of
incoming radiation at the boundaries \cite{leaver85}. A standard
technique to overcome the difficulties is the so-called Leaver's
method \cite{leaver85}, which we employ in this paper. In fact, this
seems to be the first attempt at trying to find instabilities using
this method, and it works well, as we shall see. In what follows, we
shall describe the four dimensional ($n=0$) case thoroughly, for
concreteness. The method is easily adapted to any number of
dimensions $n$, but the equations get much lengthier, so we refrain
from presenting them here. We pick units such that $M=1$. First, let
us consider the radial equation. Following Leaver, the radial
function $R$ can be expanded around the horizon as
\begin{equation}
R=(r-r_+)^{-i\sigma}(r-r_-)^{-1+i\sigma+i\mu^2(2\omega_1)^{-1}+i\omega_1}
{\rm e}^{i\omega_1 r} \sum^\infty_{k=0}b_k\left({r-r_+\over
r-r_-}\right)^k\,, \label{expansion}
\end{equation}
where $\sigma$ and $\omega_1$ are defined by
\begin{equation}
\sigma={(r_+^2+a^2)\omega-ma\over a^2+3r_+^2}\,r_+\,, \quad
\omega_1=\sqrt{\omega^2-\mu^2}\,.
\end{equation}
Here, $b_0$ is taken to be $b_0=1$, and $r_+$ and $r_-$ are,
respectively, the coordinate radius of the event and the Cauchy
horizon of the Kerr black hole, given by $r_\pm=(1\pm b)/2$, where
$b=(1-4a^2)^{1/2}$. Note that the branch of $\sqrt{z}$ has been
chosen such that $-\pi/2<arg\sqrt{z}\le\pi/2$. The expansion
coefficients $b_k$ in equation (\ref{expansion}) are determined via
the five-term recurrence relation, given by
\begin{eqnarray}
& &
\alpha_0b_1+\beta_0b_0=0\,\,,\,\,\alpha_1b_2+\beta_1b_1+\gamma_1b_0=0\,\,,\,\,
\alpha_2b_3+\beta_2b_2+\gamma_2b_1+\delta_2b_0=0\,, \nonumber \\
&&\alpha_kb_{k+1}+\beta_kb_k+\gamma_kb_{k-1}+\delta_kb_{k-2}+
\epsilon_kb_{k-3}=0\,, \ k=3,4,5,\cdots, \label{five}
\end{eqnarray}
where
\begin{eqnarray}
\alpha_k&=&(1+k)(1+k+h_0) \,\,,\,\, \beta_k=-4k^2+h_1k+o_1 \,\,,\,\,
\gamma_k=6k^2+h_2k+o_2 \,, \nonumber \\
\delta_k&=&-4k^2+h_3k+o_3 \,\,,\,\,
\epsilon_k=k^2+h_4k+o_4 \,. \nonumber \\
\label{coef1}
\end{eqnarray}
Here, the coefficients appearing in equation (\ref{coef1}) are given
by
\begin{eqnarray}
h_0&=&-ib^{-1}\{-2am+\omega(1+b)\}\,, \nonumber \\
h_1&=&-2-8iamb^{-1}+i\{4\omega b^{-1}(1+b)+
\mu^2\omega_1^{-1}+2\omega_1(1+b)\}\,, \nonumber \\
h_2&=&-i(b\omega_1)^{-1}[6\omega_1(\omega-2am)+4b^2\omega_1^2+
3b\{\mu^2+2\omega_1(\omega+\omega_1-i)\}]\,,  \\
h_3&=&10-8iamb^{-1}+i\{4\omega(1+b^{-1})+3\mu^2\omega_1^{-1}+2\omega_1(3+b)\}
\,, \nonumber \\
h_4&=&-i(b\omega_1)^{-1}[\omega_1(\omega-2am)+b\{\mu^2+\omega_1(\omega+2\omega_1-4i)\}]
\,, \nonumber
\end{eqnarray}
%
and
\begin{eqnarray}
o_1&=&(4b\omega_1^2)^{-1}[b^3\omega_1^4+2(2am-\omega)\{-2\omega^2(i+\omega+\omega_1)+
\mu^2(2i+2\omega+\omega_1)\}\nonumber \\
&&-2b^2\omega_1^2\{\mu^2-2(\omega^2+i\omega_1(i+\omega))\}\nonumber \\
&&+b\{\mu^4+\mu^2(4+4A_{lm}-4i\omega-8\omega^2-2i\omega_1+8am\omega_1-6\omega\omega_1)
\nonumber \\
&&{\quad}+\omega^2(-4-4A_{lm}+4i\omega+7\omega^2+4i\omega_1-8am\omega_1+8\omega\omega_1)\}]
\,, \nonumber \\
o_2&=&(4b\omega_1^2)^{-1}[-2b^3\omega_1^4-4b^2\{\mu^4-\mu^2(3\omega^2+2i\omega_1+2\omega\omega_1)+
2(\omega^4+i\omega^2\omega_1+\omega^3\omega_1-2i\omega_1^3)\}
\nonumber \\
&&-6(2am-\omega)\{\mu^2(2i+2\omega+\omega_1)-2(\omega^3-2i\omega_1^2+
\omega^2(i+\omega_1))\}
\nonumber \\
&&+b\{-3\mu^4-2\mu^2(6+4A_{lm}-6i\omega-10\omega^2-
10i\omega_1+8am\omega_1-7\omega\omega_1)
\nonumber \\
&&{\quad}+2(-9\omega^4+12i\omega\omega_1^2+
12i\omega_1^3-2\omega^3(3i+5\omega_1)+\omega^2(6+4A_{lm}-8i\omega_1+8am\omega_1))\}]
\,,  \\
o_3&=&(4b\omega_1^2)^{-1}[b^3\omega_1^4+b\{3\mu^4+15\omega^4-32i\omega\omega_1^2-
16\omega_1^2(1+3i\omega_1)+4\omega^3(3i+4\omega_1)
\nonumber \\
&&-4\omega^2(3+A_{lm}-5i\omega_1+2am\omega_1)+2\mu^2(6+2A_{lm}-6i\omega-8\omega^2-
17i\omega_1+4am\omega_1-5\omega\omega_1)\}
\nonumber \\
&&+2b^2\{\mu^4-\mu^2(3\omega^2+2i\omega_1+2\omega\omega_1)
+2(\omega^4+i\omega^2\omega_1+\omega^3\omega_1-4i\omega_1^3)\}
\nonumber \\
&&+2(2am-\omega)\{3\mu^2(2i+2\omega+\omega_1)-
2(3\omega^3-8i\omega_1^2+3\omega^2(i+\omega_1))\}]
\,, \nonumber \\
o_4&=&(4b\omega_1^2)^{-1}[-2(2am-\omega)\{\mu^2(2i+2\omega+\omega_1)-
2(\omega^3-3i\omega_1^2+\omega^2(i+\omega_1))\}
\nonumber \\
&&+b\{-\mu^4-4(\omega^4+\omega^2(-1+2i\omega_1)-
3i\omega\omega_1^2-3\omega_1^2(1+2i\omega_1)+\omega^3(i+\omega_1))
\nonumber \\
&&+2\mu^2(-2+2\omega^2+8i\omega_1+\omega(2i+\omega_1))\}] \,,
\nonumber
\end{eqnarray}
Using Gaussian elimination twice, one can reduce the five-term
recurrence relations (\ref{five}) to three-term recurrence
relations, which can be written as
\be
\alpha'_0b_1+\beta'_0b_0=0\,,\,\,\,\,\,\,\alpha'_kb_{k+1}+\beta'_kb_k+\gamma'_kb_{k-1}\,,
\ k=1,2,\cdots, \label{three} \ee
Since the Gaussian elimination is straightforward, we do not show an
explicit procedure for the Gaussian elimination (see, e.g.,
\cite{leaver90}). An eigenfunction satisfying the quasinormal mode
boundary conditions behaves at the event horizon and infinity as
\begin{equation}
\Psi\sim\left\{
\begin{array}{ll}
(r-r_+)^{-i\sigma} & {\rm as}\ r\rightarrow r_+ \,, \\
r^{-1+i\mu^2(2\omega_1)^{-1}+i\omega_1} {\rm e}^{i\omega_1 r} & {\rm
as}\ r\rightarrow \infty\,.
\end{array}
\right. \label{bc}
\end{equation}
Therefore, one can see that the expanded wave function
(\ref{expansion}) satisfies the quasinormal mode boundary conditions
if the expansion in (\ref{expansion}) converges at spatial infinity.
This convergence condition for the expansion (\ref{expansion}),
namely the quasinormal mode conditions, can be written in terms of
the continued fraction as \cite{Gu67,leaver85}
\begin{eqnarray}
\beta'_0-{\alpha'_0\gamma'_1\over\beta'_1-}
{\alpha'_1\gamma'_2\over\beta'_2-}{\alpha'_2\gamma'_3\over\beta'_3-}
...\equiv \beta'_0-\frac{\alpha'_0\gamma'_1}{\beta'_1-
\frac{\alpha'_1\gamma'_2}{\beta'_2-\frac{\alpha'_2\gamma'_3}{\beta'_3-...}}}
=0 \,, \label{a-eq}
\end{eqnarray}
where the first equality is a notational definition commonly used in
the literature for infinite continued fractions. Here, we shall
adopt this convention.

Next, we turn ourself to the angular equation. In order to determine
the separation constant $A_{lm}$, a similar technique to that used
for the radial equation can be applied. Imposing regularity on the
symmetry axis, we can expand the angular function as
\begin{eqnarray}
S={\rm e}^{a\omega_1u}(1-u^2)^{|m|\over
2}\sum_{k=0}^{\infty}d_k(1+u)^k\,, \label{ang-exp}
\end{eqnarray}
where $u=\cos\theta$, and $d_0=1$. The expansion coefficients $d_k$
in equation (\ref{ang-exp}) are determined via the three-term
recurrence relation, given by
\be
\tilde\alpha_0d_1+\tilde\beta_0d_0=0\,\,,\,\,\tilde\alpha_kd_{k+1}+\tilde\beta_kd_k+\tilde\gamma_kd_{k-1}=0\,,
\ k=1,2,3,\cdots, \label{ang-three} \ee
where
\begin{eqnarray}
\tilde\alpha_k&=&-2(1+k)(k+|m|+1) \,, \nonumber \\
\tilde\beta_k&=&k(k-1)+2k(|m|+1-2a\omega_1)-(2a\omega_1-|m|)(|m|+1)-(a^2\omega_1^2+A_{lm}) \,, \label{coef2} \\
\tilde\gamma_k&=&2a\omega_1(k+|m|) \,. \nonumber
\end{eqnarray}
Note that in the four dimensional case, we have to set $j=0$ in
equations (2.5) and (2.6). If the expansion in equation
(\ref{ang-exp}) converges for $-1\le u\le 1$, the angular function
satisfies the regularity conditions at $u=\pm 1$. This condition is
equivalent to the continued fraction equation, given by
\cite{Gu67,leaver85}
\begin{eqnarray}
\tilde\beta_0-{\tilde\alpha_0\tilde\gamma_1\over\tilde\beta_1-}
{\tilde\alpha_1\tilde\gamma_2\over\tilde\beta_2-}
{\tilde\alpha_2\tilde\gamma_3\over\tilde\beta_3-} ... =0 \,,
\label{aa-eq}
\end{eqnarray}

Now that we have two continued fraction equations of the frequency
$\omega$ and the separation constant $A_{lm}$ (\ref{a-eq}) and
(\ref{aa-eq}), we can obtain the frequency $\omega$ and the
separation constant $A_{lm}$ solving equations (\ref{a-eq}) and
(\ref{aa-eq}) simultaneously. These coupled algebraic equations
(\ref{a-eq}) and (\ref{aa-eq}) can be solved numerically (see, e.g.,
\cite{n-recipes,nollert93}).
\subsection{The instability for the ${\rm d}=4$ case}
\begin{figure}
\centerline{\mbox{ \psfig{figure=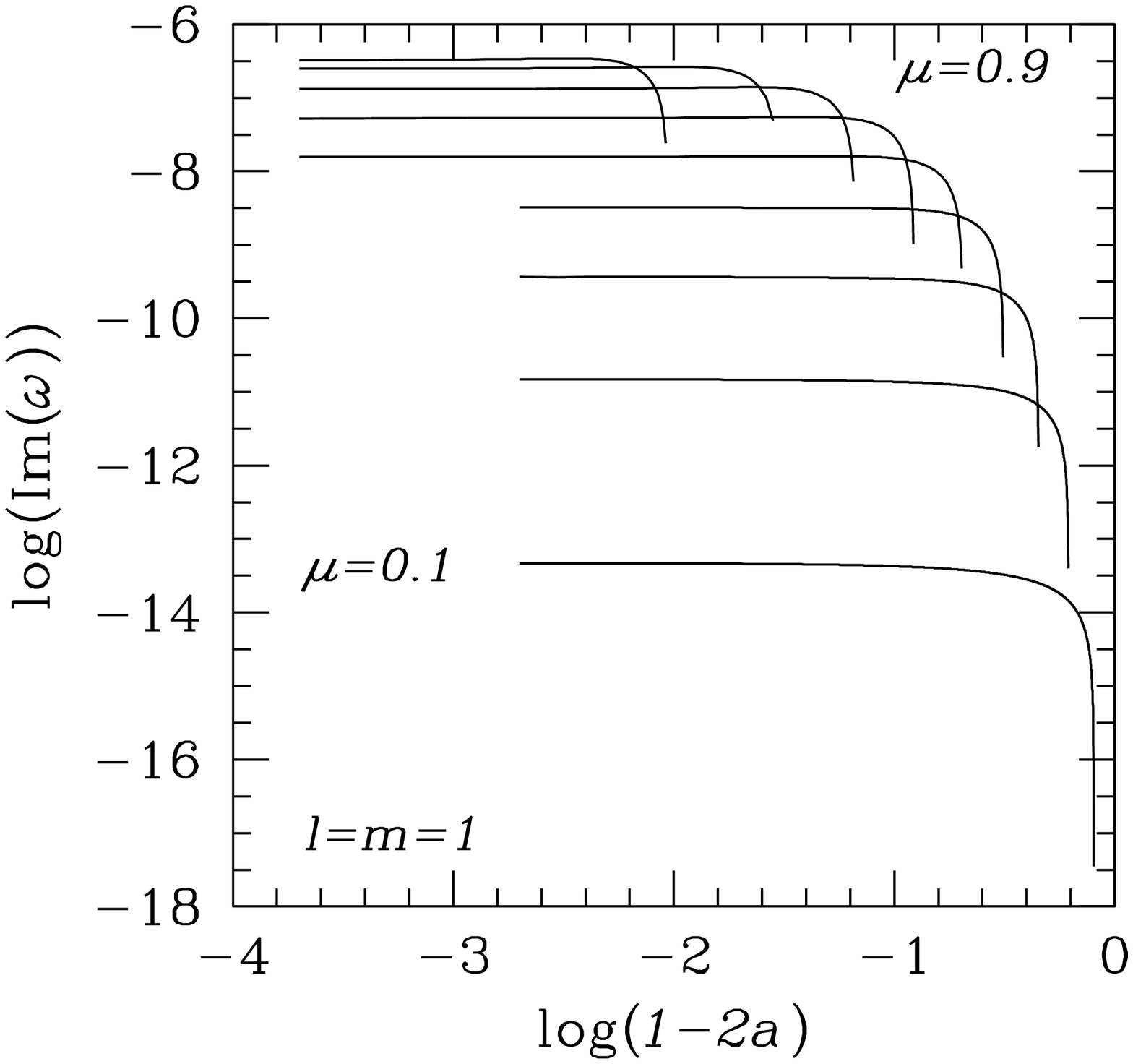,angle=0,width=6cm}
\psfig{figure=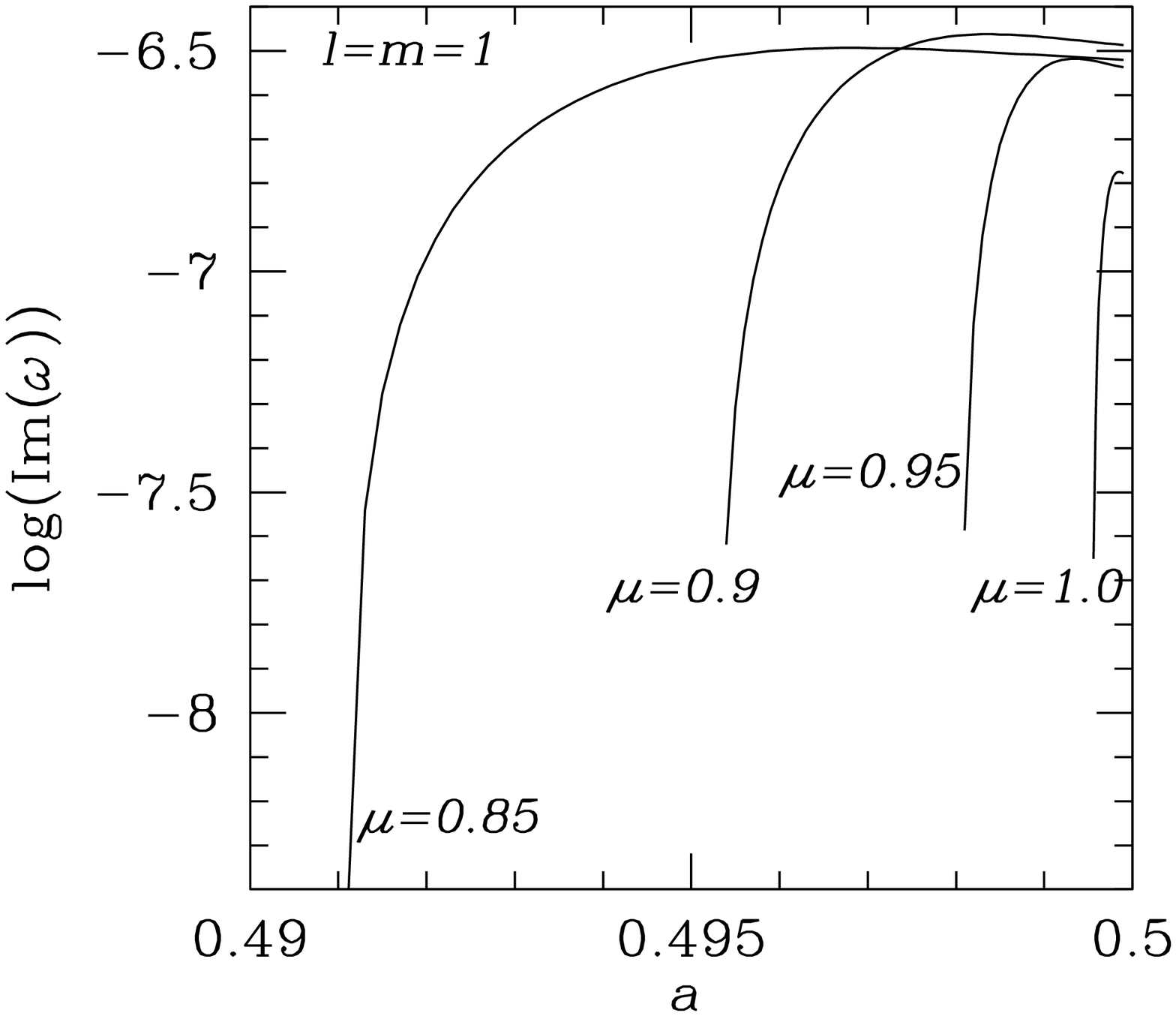,angle=0,width=6cm}}} \centerline{\mbox{
}}
\caption{Details for the instability in the four dimensional case.
We have set $M=1$, so the rotation parameter $a$ varies between $0$
(Schwarzschild limit) and $1/2$ (extremal limit). In the left panel
we plot the imaginary part of the unstable modes as a function of
$a$ for nine values of $\mu$, and for $l=m=1$. The same quantities
are plotted in the right panel, but now we focus on the maximum
instability region, for values of $\mu$ near unity. The instability
is stronger for larger values of $a$ and for $\mu \sim 1$. In fact,
our results suggest that the imaginary part of the unstable modes
attains its maximum value, $10^{-6}\times{\rm Re}[\omega]$ for
$\mu=0.9$ and $a=0.497$. Notice finally that as one decreases $a$
(going right on the $x$ axis in the left panel), there is a critical
$a$ below which there is no instability. Although there are an
infinity of unstable modes, we only show here the most unstable
one.} \label{fig:superradiantinstability1}
\end{figure}

\begin{figure}
\centerline{\mbox{
\psfig{figure=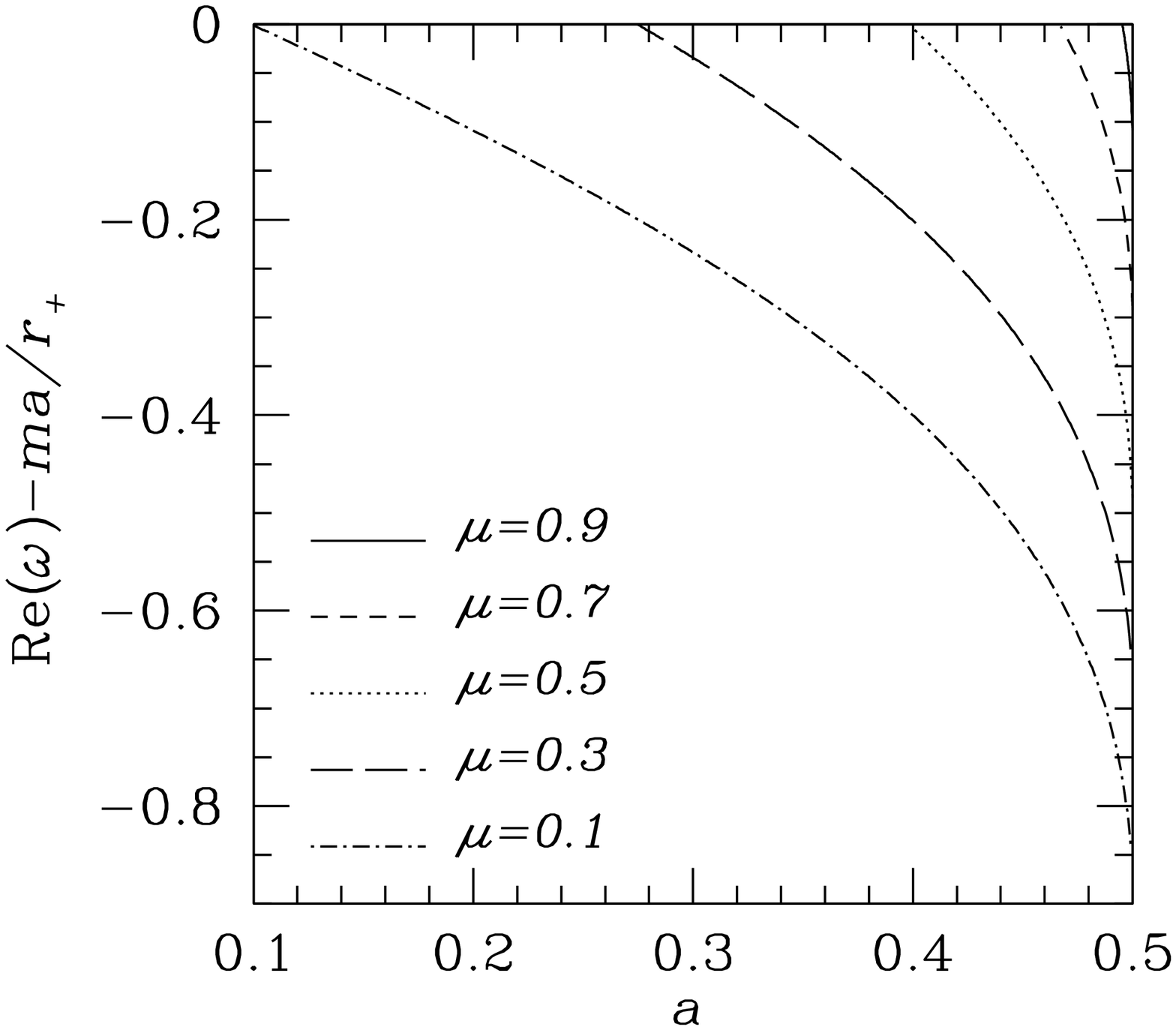,angle=0,width=6cm}
\psfig{figure=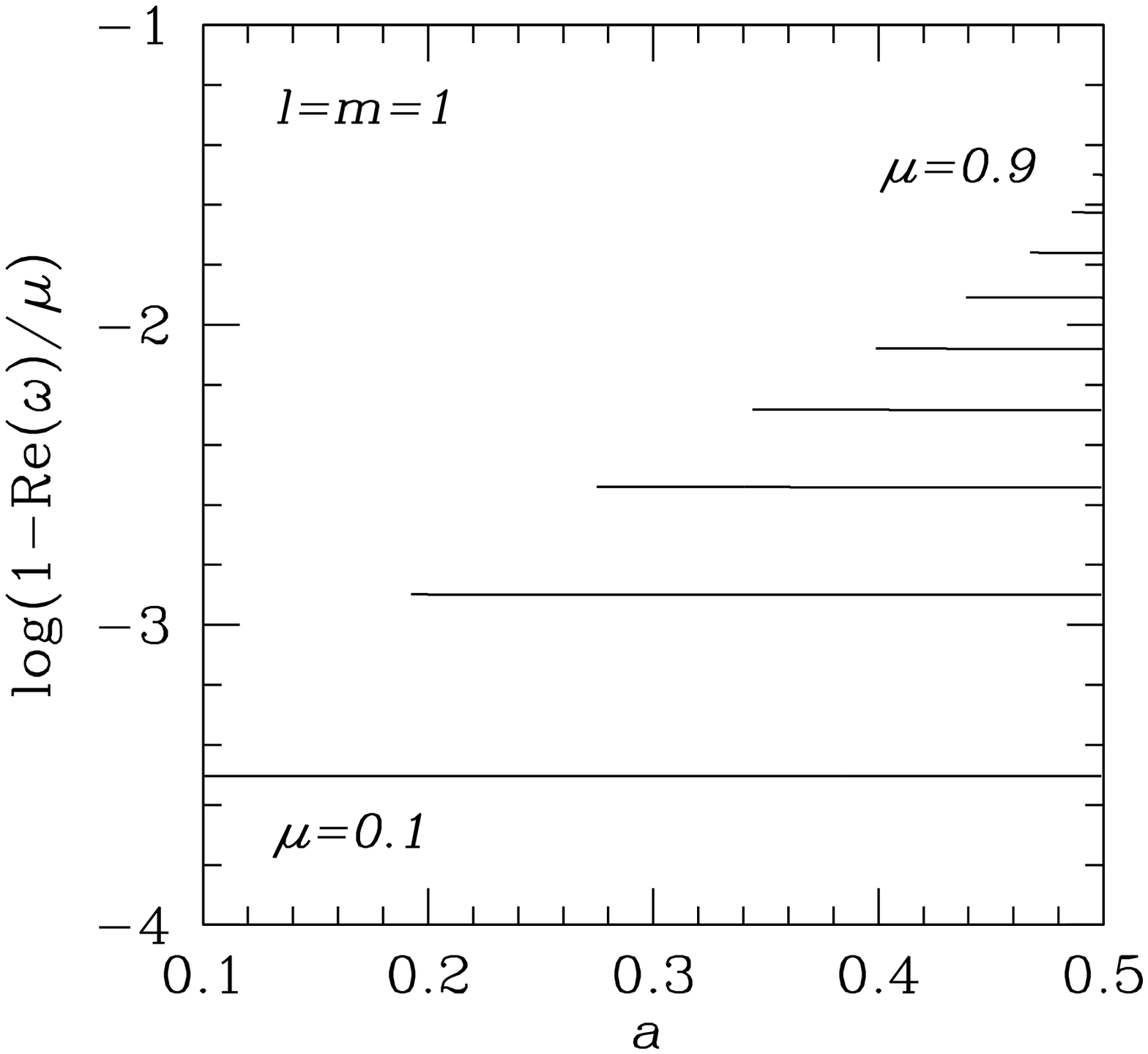,angle=0,width=6cm} }}
\caption{More details of the instability in the four dimensional
case. Here we plot the superradiant factor ${\rm
Re}[\omega]-\frac{ma}{a^2+r_+^2}$ as a function of $a$ for $l=m=1$
and for several values of $\mu$. Again, we have set $M=1$ and so the
superradiant factor is, in four dimensions, ${\rm
Re}[\omega]-\frac{ma}{r_+}$. the instability disappears for $a$
below a certain critical point, which is given by setting the
superradiant factor to zero.
 On the right panel we make the same plot
but on a logarithmic scale, for better visualization. Notice that
the superradiant factor is negative, as it should for the
instability to be triggered.} \label{fig:superradiantinstability2}
\end{figure}

Some numerical results for ${\rm d=4}$ were given by Furuhashi and
Nambu \cite{massivescalar} and also by Strafuss and Khanna
\cite{massivescalar}, in the context of massive field instability of
the Kerr metric which as we saw, translate immediately to our
situation. Nevertheless, we have performed an independent extensive
numerical search, which allow us to have a more complete picture and
understanding of the situation. Our numerical results were obtained
using the technique just described {\it supra}, and are summarized
in Figures
\ref{fig:superradiantinstability1}-\ref{fig:superradiantinstability2}.
There is more than one frequency with a positive imaginary part,
corresponding to unstable modes. We only show the most relevant one
which is the one having a larger imaginary part, and therefore
corresponds to the most unstable mode.

In Figure \ref{fig:superradiantinstability1} we plot the imaginary
part of the (unstable) mode as a function of $a$ for several values
of $\mu$ and for $l=m=1$. The instability timescale is given by
$\tau \sim \frac{1}{{\rm Im}[\omega]}$. We can see that:

(i) The instability gets stronger (the typical timescale decreases)
as the rotation $a$ increases. This is expected, since we know that
the superradiant amplification gets stronger as $a$ increases
\cite{cardosoetal,ampfact}.

(ii) The instability also gets stronger as $\mu$ increases. This has
a very simple explanation: the depth of the potential well in Figure
\ref{fig:potentialwell} is larger for larger values of $\mu$, and
therefore the waves get more efficiently trapped. Our results
suggest that the imaginary part of the unstable modes attains its
maximum value, $10^{-6}\times{\rm Re}[\omega]$ for $l=m=1$,
$\mu=0.9$ and $a=0.497$.

(iii) For a fixed $\mu$ and as one decreases $a$ the unstable modes
disappear below a certain critical value of $a$. This is because the
superradiance condition ${\rm Re}[\omega]-\frac{ma}{a^2+r_+^2}<0$ is
not satisfied for very low rotation parameter $a$. This is best seen
in Figure \ref{fig:superradiantinstability2}.

In Figure \ref{fig:superradiantinstability2} we plot the real part
of the (unstable) mode as a function of $a$ for several values of
$\mu$ and for $l=m=1$. Two important features are the following:

(i) As shown in both plots of Figure
\ref{fig:superradiantinstability2}, the superradiant factor ${\rm
Re}[\omega]-\frac{ma}{a^2+r_+^2}$ is negative, and one is therefore
in the superradiant regime. Thus, this is indeed a superradiant
instability.

(ii) The instability occurs only for $\omega<\mu$. Otherwise, the
waves would not be trapped in the potential well, but they would
rather escape to infinity, and the perturbation would be damped.

Detweiler \cite{massivescalar} finds (in four dimensions), in the
limit of small $\mu$, that the characteristic frequencies of the
unstable modes are, in our units, \be \omega \sim \mu +\ii \mu
\frac{a}{M}\frac{(\mu M)^8}{3072}\,.\label{detform} \ee Our
numerical results fit this prediction very well. For near-extremal
black holes, and $\mu \sim 0.1$ Detweiler's formula (\ref{detform})
predicts $\omega \sim 0.1+10^{-13}\ii$, which checks very well with
our numerics (see Figure \ref{fig:superradiantinstability1} and
\ref{fig:superradiantinstability2}).

Although of no direct interest for this work, we found also stable
modes. Our results for stable modes will be published elsewhere
\cite{shijun}, and compared with the existing analytical ones
\cite{will}.
\subsection{No instability for higher dimensional rotating Kerr-like black branes}
We have also performed a numerical search for the instability in the
${\rm d}=5\,,6\,,7$ case. We found no trace of unstable modes, and
we justify this in a rigorous analysis presented in Appendix
\ref{sec:A0}. We now turn to explain why higher dimensional
Kerr-like branes should be stable against this particular mechanism.

We saw that the instability arises because of superradiantly
amplified trapped modes, in the potential well. But does a well
exist for general ${\rm d}$? It doesn't, and to understand this, we
have to look at the asymptotic behavior of the effective potential
(\ref{effV}). If the derivative of this potential is positive near
spatial infinity, we are guaranteed to have a well, and therefore an
instability. If the derivative is negative, the modes should all be
stable. Near infinity, the dominant terms in the effective potential
(\ref{effV}) are \be V' \sim -\frac{2}{r^3} (A_{lm}+a^2\mu
^2+\frac{n}{2}+\frac{n^2}{4}-a^2\omega ^2)+(n+1)\mu ^2M r^{-2-n} \,,
\ee where we have already substituted for the separation constant
$\lambda =A_{lm}-2am\omega+a^2\omega ^2$. It is immediately apparent
that the four dimensional case is a special one: if $n=0$, the
dominant term in the derivative is $(n+1)\mu ^2M r^{-2-n}$, which is
positive and we therefore are guaranteed to have a bound state.
Thus, this case should be unstable, and it is, as we just described
in the previous subsection.

When $n>0$ the other terms dominate. In fact, for $n>1$ they are
positive. For $n>1$ the dominant terms are \be V' \sim
-\frac{2}{r^3} (A_{lm}+a^2\mu ^2+\frac{n}{2}+\frac{n^2}{4}-a^2\omega
^2)\,. \ee Since $\omega<\mu$, this is negative (the separation
constant $A_{lm}$ can be shown to be positive). Thus, for ${\rm
d}>5$ there is no potential well, no bound states and therefore no
instability, even though there is still superradiance. The situation
for $n=1$ is not as clear, because there is the extra term $(n+1)\mu
^2M r^{-2-n}$. In principle it should be possible to have, for
certain very specific parameters, a potential well. But to do that,
one would have to require that $\mu$ be very large, and this makes
it very hard to study the problem numerically (the imaginary part is
expected to be extremely small in this regime, as shown by Zouros
and Eardley \cite{massivescalar}, and this prevents any numerical
treatment).

There is another, perhaps more physical, explanation of the fact
that higher dimensional rotating black branes should be stable. We
know that the a massless field propagating in the background of
these rotating black branes is equivalent to a massive field
propagating in the background of a Myers-Perry black hole. Imagine
now a wavepacket of the massive field in a distant circular orbit.
In the four-dimensional situation, the gravitational force binds the
field and keeps it from escaping or radiating away to infinity. But
at the event horizon some of the field goes down the black hole, and
if the frequency of the field is in the superradiant region then the
field is amplified. Hence the field is amplified at the event
horizon while being bound away from infinity. We should thus have an
instability, as we do. However, for ${\rm d}>4$, there are no stable
orbits \cite{orbits}, and thus the field escapes to infinity, no
instability is triggered. This complements the previous view point.
Escaping to infinity is equivalent to not having a potential well to
bound the field.

\section{Acoustic black branes}
In 1981 Unruh \cite{unruh} introduced the notion of ``dumb holes'',
which is an example of analogue black holes. While not carrying
information about Einstein's equations, the analogue black holes
devised by Unruh do have a very important feature that defines black
holes: the existence of an event horizon. The basic idea behind
these analogue acoustic black holes is very simple: consider a fluid
moving with a space-dependent velocity, for example water flowing
throw a variable-section tube. Suppose the water flows in the
direction where the tube gets narrower. Then the fluid velocity
increases downstream, and there will be a point where the fluid
velocity exceeds the local sound velocity, in a certain frame. At
this point, in that frame, we get the equivalent of an apparent
horizon for sound waves. In fact, no (sonic) information generated
downstream of this point can ever reach upstream (for the velocity
of any perturbation is always directed downstream, as a simple
velocity addition shows). This is the acoustic analogue of a black
hole, or a {\it dumb hole}. We refer the reader to \cite{visser} for
a review of these objects.
\subsection{The wave equation for a general space-dependent density}
We can also model acoustic black branes easily, and we can thus look
for the instability dealt with here in the laboratory. For
simplicity, let us consider the following black brane. First build a
$2+1$ dimensional acoustic black hole \cite{visser}, modelling a
draining bathtub. Consider a fluid having (background) density
$\rho(x)$. Assume the fluid to be locally irrotational (vorticity
free), barotropic and inviscid. From the equation of continuity, the
radial component of the fluid velocity satisfies $\rho v^r\sim 1/r$.
Irrotationality implies that the tangential component of the
velocity satisfies $v^\theta\sim 1/r$. We shall keep both $\rho$ and
the sound velocity $c$ as position-dependent quantities, thus
generalizing the treatment in \cite{visser,cardosoanalog1}. We then
have \beq v^r=\frac{A}{\rho r} \,\,,\,\,
v^{\theta}=\frac{B}{r}\,,\eeq where $A\,,\,B$ are constants. The
acoustic metric describing the propagation of sound waves in this
``draining bathtub'' fluid flow is \cite{visser}:

\begin{equation}
ds^2= -\left (c^2-\frac{A^2/\rho ^2+B^2}{r^2} \right
)dt^2+\frac{2A}{\rho r}drdt-2Bd\phi dt+dr^2+ r^2d\phi^2.
\label{metric1}
\end{equation}
It is however better to work with a more transparent metric. Some
physical properties of this draining bathtub metric are more
apparent if we cast the metric in a Kerr-like form performing the
following coordinate transformation (see
\cite{cardosoanalog1,basak}):
\begin{eqnarray}
dt&\rightarrow& d\tilde{t}=dt-\frac{Ar/\rho}{r^2c^2-A^2/\rho^2}dr\\
d\phi&\rightarrow&
d\tilde{\phi}=d\phi-\frac{BA/\rho}{r(r^2c^2-A^2/\rho ^2)}dr\,,
\label{coordtransf}
\end{eqnarray}
Then the effective metric takes the form
\be ds^2= -\left(1-\frac{A^2/\rho ^2+B^2}{c^2r^2} \right)c^2
d\tilde{t}^2 + \left(1-\frac{A^2}{\rho ^2c^2r^2} \right )^{-1}dr^2
-2B d\tilde{\phi}d\tilde{t}+r^2d\tilde{\phi}^2\,.\label{metric2} \ee
As explained in \cite{cardosoanalog1}, this metric and the Kerr
metric differ in an important aspect, in that whereas the rotation
for the Kerr black hole is bounded from above, here it is not, at
least in principle. Thus, $B$ could be as large as desired.

Now, we are free to add an extra dimension $z$ to this effective
geometry, as explained by Visser \cite{visser}, resulting in the
following black brane geometry

\beq ds^2= -\left(1-\frac{A^2/\rho ^2+B^2}{c^2r^2} \right)c^2
d\tilde{t}^2 + \left(1-\frac{A^2}{\rho ^2c^2r^2} \right )^{-1}dr^2
-2B d\tilde{\phi}d\tilde{t}+r^2d\tilde{\phi}^2+dz^2\,.
\label{metric3} \eeq The propagation of a sound wave in a barotropic
inviscid fluid with irrotational flow, which is assumed to be the
case, is described by the Klein-Gordon equation
$\nabla_{\mu}\nabla^{\mu}\Phi=0$ for a massless field $\Psi$ in a
Lorentzian acoustic geometry. Separating variables by the
substitution
\be \Phi(\tilde{t},r,\tilde{\phi})=
 \sqrt{r} \Psi(r)e^{i (\mu z + m\tilde{\phi}-\omega\tilde{t})}\,,
\ee
implies that $\Psi(r)$ obeys the wave equation

\be \frac{d^2 \Psi}{dr_* ^2}+\left( (\omega -\frac{Bm}{r^2})^2 -V
\right )\Psi=0\,.\ee

Here \be V= f\left (\mu
^2c+\frac{4m^2c-c}{4r^2}+\frac{5A^2}{4cr^4\rho
^2}+\frac{c'}{2r}+\frac{A^2}{2cr^3\rho ^2}(\frac{c'}{c}+\frac{2\rho
'}{\rho}) \right )\,,\ee and the tortoise coordinate $r_*$ is
defined as \be \frac{dr}{dr_*}=c(1-\frac{A^2}{c^2\rho ^2 r^2})\equiv
f\,.\ee Notice that for constant $\rho\,,\, c$ one recovers the
equations in \cite{cardosoanalog1,basak}, as one should. Now, it is
quite easy to present an example of background flow for which the
instability is triggered: take for instance a flow for which $\rho$
is almost constant at infinity (almost means that it asymptotes to a
constant value more rapidly than the sound velocity). Assume also
that, near infinity, $c=c_1+\frac{c_2}{r}$. Then, we get that near
infinity the effective potential behaves as \be
\frac{2c_1c_2k^2}{r}\,.\ee For this to have a positive derivative,
one requires $c_2<0$ ($c_1$ must be positive, as it is the
asymptotic value of the sound velocity). We thus have one example of
flow for which the instability is active. There are many others, of
course, and there are also instances for which the system is stable.
But the important point here, is that we {\it can} build this
effective geometry in the lab, and thus observe this instability.
\section{The endpoint of the instability}
We have shown that many rotating black objects are unstable against
the instability displayed here. We have also shown that, for a fixed
mass term $\mu$, the instability disappears for a rotation parameter
$a$ below a certain value, for which the superradiant factor ${\rm
Re}[\omega]-\frac{ma}{a^2+r_+^2}$ is zero. Now, as the unstable mode
grows, rotational energy is being extracted from the black brane,
and thus its rotation decreases (the same phenomenon happens for the
black hole bomb \cite{cardosoetal}). This means that eventually the
instability stops growing, when the rotation is in the critical
point, for which the superradiant factor is zero. It looks very
tempting and reasonable to assume that the system will settle down
to a low-rotation black brane plus some radiation around it.
Eventually, all this radiation will escape to infinity. The total
angular momentum of the system is conserved, so this radiation
carries the extra angular momentum. This is also clear from the fact
that only co-rotating modes are unstable. This is one possibility.
But bearing in mind that these are bound modes, it seems plausible
that one other thing could happen, before the radiation escapes to
infinity: since we are dealing with higher dimensional objects, the
end product could be an object with two or more components for the
angular momentum, that is to say, there could be a transfer of
angular momentum from one dimension to the other.

\section{Conclusions}
We have shown that there is a large class of extended rotating black
objects which are unstable against {\it massless} field
perturbations. The instability is caused by superradiant
amplification of waves trapped inside the potential well, or from a
different view-point, it is caused by superradiant amplification of
wave packets bound by the gravitational force. This mechanism is
very similar to other superradiant mechanisms in the presence of
black holes \cite{teupress,cardosoetal,massivescalar}, or other
rotating objects such as stars \cite{superstars}. The important
point here is that the extra dimensions work as an effective mass
for any massless field. Thus, the evolution of a massless scalar
field in, for example, the black brane (\ref{blackkerrD}) is
equivalent to the evolution of a massive field in the background of
a Myers-Perry rotating black hole. In four dimensions, the black
hole displays superradiance and has bound states due to the mass
term, thus the geometry is unstable. In higher dimensions, the
spacetimes dealt with here have no stable circular geodesics (the
mass term does not give rise to a potential well in the wave
description) and therefore these spacetimes are stable. This simple
reasoning implies that general  four dimensional extended rotating
black objects (more general than the one described by
(\ref{blackkerr})) will be unstable. It also seems to imply that
general higher dimensional rotating black objects are stable.
Moreover, following Zel'dovich \cite{zelmisnerunruhbekenstein}, it
is known that not only the Kerr geometry, but any rotating absorbing
body for that matter displays superradiance. Thus, this instability
appears also in analogue black hole models, as we have shown.

Here, we have derived the instability timescale for a scalar field,
and not for geometry (metric) perturbations, since Teukolsky's
formalism for higher dimensional rotating objects is still not
available. Still, the argument presented above makes it clear that
the instability should be present for metric perturbations as well.
Indeed, the presence of the potential well, in four dimensions, is
due only to the mass term, and not to the geometry properties
themselves, and since metric modes also scatter superradiantly
\cite{ampfact} then everything we discussed here translates
immediately to metric perturbations.
 We also expect that the instability will be stronger for
metric modes, because of the following simple reasoning.
Superradiance is responsible for this instability, and thus the
larger the superradiant effects, the stronger the instability. Now,
we know that in the four dimensional Kerr geometry scalar fields
have a maximum superradiant amplification factor of about $2\%$,
whereas gravitational modes have maximum superradiant amplification
factor of about $138 \%$ \cite{ampfact}. Thus, we expect the
instability timescale to be almost two orders of magnitude smaller
for gravitational modes, which means that the instability is two
orders of magnitude more effective for metric modes. For general
black branes, the Gregory-Laflamme instability seems to be stronger
than the one displayed here, but it is known that certain extremal
solutions should not exhibit the Gregory-Laflamme instability
\cite{reall}, whereas the instability dealt with here should go all
the way to extremality. So eventually it takes over the
Gregory-Laflamme instability. Moreover, recent studies
\cite{maartens} seem to indicate that black strings in a
Randall-Sundrum inspired 2-brane model do not exhibit the
Gregory-Laflamme instability, but they should be unstable against
this mechanism.

It seems plausible to assume that the instability will keep growing
until the energy and angular momentum content of the field
approaches that of the black brane, when back-reaction effects
become important. The rotating brane will then begin to spin down,
and gravitational and scalar radiation goes off to infinity carrying
energy and angular momentum. The system will probably be asymptotic
to a static, or very slowly rotating, final state consisting of a
non-rotating black p-brane and some outgoing radiation at infinity.
The end product could also be an object with two or more components
for the angular momentum, that is to say, there could be a transfer
of angular momentum from one dimension to the other.

\vskip1.5cm
\section*{Acknowledgements}
We would like to thank Jim Meyer for a critical reading of the
manuscript. V. C. acknowledges financial support from Funda\c c\~ao
para a Ci\^encia e Tecnologia (FCT) - Portugal through grant
SFRH/BPD/2004. S.Y. is supported by the Grant-in-Aid for the 21st
Century COE ``Holistic Research and Education Center for Physics of
Self-organization Systems'' from the ministry of Education, Science,
Sports, Technology, and Culture of Japan.
\appendix
\section{\label{sec:A0}Factorized potential analysis}
We now present a full description of our search for bound states in
general ($4+n$)-dimensions. We conclude that no unstable bound
states exist for $n\geq 5$. The massive scalar field equation in 4+n
dimensional spacetime is given by
\begin{equation}
\Delta r^{-n}{d\over dr}\left(\Delta r^n {dR\over dr}\right)+V_0
R=0\,,
\end{equation}
where $V_0$ is the effective potential, given by
\begin{equation}
V_0=\{\omega(r^2+a^2)-am\}^2-\Delta\left\{\mu^2r^2+A_{lm}-2m\omega
a+ \omega^2 a^2+ {j(j+n-1)a^2\over r^2}\right\}\,. \label{master-eq}
\end{equation}
For simplicity, let us assume $A_{lm}$ to be $A_{lm}=l(l+1)$. Note
that this assumption is only valid in the limit of
$a^2(\omega^2-\mu^2)\rightarrow 0$. Introducing a new variable $r_*$
and a normalized function $\Phi$, defined as
\begin{equation}
dr_*={r^2+a^2\over\Delta}\,dr\,,\quad R=\{(r^2+a^2)r^n\}^{-{1\over
2}}\Phi\,,
\end{equation}
we can reduce equation (\ref{master-eq}) to
\begin{equation}
{d^2\Phi\over dr_*^2} +V \Phi=0\,,
\end{equation}
where
\begin{eqnarray}
V={V_0\over (r^2+a^2)^2}-G^2-{dG\over dr_*}\,,\quad G={d\over
dr_*}\log[(r^2+a^2) r^n]^{1\over 2}\,.
\end{eqnarray}
Since $V$ is, due to the simplified $A_{lm}$, a quadratic function
of $\omega$, it can be factorized as
\begin{eqnarray}
V=\alpha(\omega-V_+)(\omega-V_-)\,,
\end{eqnarray}
where $V_+\ge V_-$ and $\alpha$ is a positive function, given by
\begin{eqnarray}
\alpha&=&{(r^2+a^2)^2-\Delta a^2\over (r^2+a^2)^2} \nonumber \\
      &=&{(r^2+a^2)^2-a^2(r^2+a^2-r^{1-n})\over (r^2+a^2)^2} \nonumber \\
      &=&{(r^2+a^2)r^2+a^2 r^{1-n}\over (r^2+a^2)^2} > 0\,.
\end{eqnarray}
Note that $\alpha$ also satisfies
\begin{eqnarray}
\lim_{r\rightarrow r_+}\alpha=\lim_{r\rightarrow\infty}\alpha=1\,.
\end{eqnarray}
The factorized potentials $V_\pm$ have the following properties:
\begin{eqnarray}
\lim_{r\rightarrow r_+}V_\pm={ma\over r^2_++a^2}=m\Omega\,, \quad
\lim_{r\rightarrow\infty}V_\pm=\pm\mu\,,
\end{eqnarray}
where $\Omega$ is the rotation frequency of the black hole at the
horizon. Note that $V_+=V_-$ is satisfied at the horizon.

\begin{figure}
\centerline{ \mbox{\psfig{figure=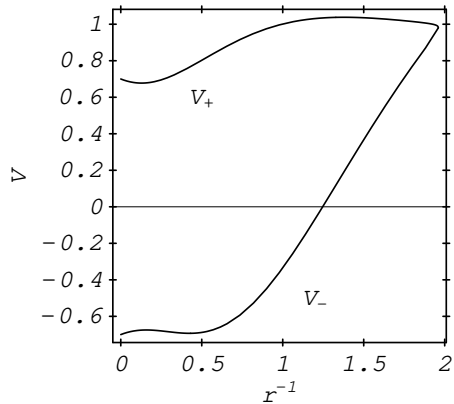,angle=0,width=6cm}}}
\caption{The factorized potentials $V_\pm$ as a function of
$r^{-1}$. The potentials are characterized by $n=0$, $l=m=1$, $j=0$,
$\mu=0.7$, and $a=0.4999$.\label{fig:facpot1}}
\end{figure}

The factorized potentials are quite useful to see how a solution
behaves when the frequency $\omega$ is given. First, let us consider
the four dimensional case, i.e. the n=0 case. Taking the case of
$l=m=1$, $j=0$, $\mu=0.7$, and $a=0.4999$ as an example, we plot the
typical behavior of the factorized potentials for the four
dimensional case in Figure \ref{fig:facpot1}.
Since $\alpha$ is positive definite, the field $\Phi$ has
propagative (evanescent) character when $\omega>V_+$ or $\omega<V_-$
($V_-<\omega<V_+$) because $\Phi^{-1}d^2\Phi/dr_*^2<0$
($\Phi^{-1}d^2\Phi/dr_*^2>0$) there. In other words, the region
where $\omega>V_+$ or $\omega<V_-$ ($V_-<\omega<V_+$) is a
classically allowed (forbidden) one. By a factorized potential
diagram, we can then see where is the propagative or evanescent zone
once the frequency $\omega$ is given.

\begin{figure}
\centerline{\mbox{\psfig{figure=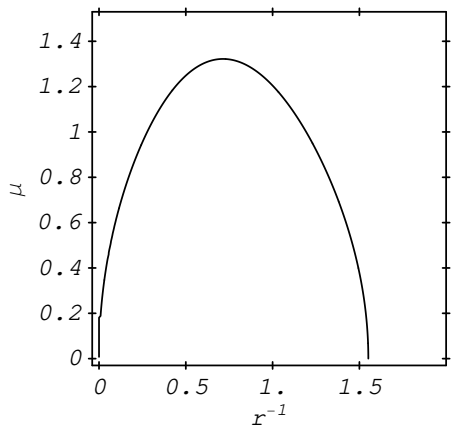,angle=0,width=6cm}}}
\caption{The curve of $dV_+/dr(r^{-1},\mu)=0$ for the parameter
$n=0$, $l=m=1$, $j=0$, and $a=0.4999$.\label{fig:facpotmais}}
\end{figure}

Next, let us consider bound states for our problem. To have a bound
state, we need a propagative zone between evanescent zones. That is
to say, unless $V_+$ ($V_-$) does have a local minimum (maximum),
there are no bound state in the potential. We can therefore expect
from the behavior of $dV_{\pm}/dr$ whether bound states are possible
or not. In Figure \ref{fig:facpotmais}, we display the $dV_{+}/dr=0$
curve on the $r^{-1}$ -- $\mu$ plan for the case of $l=m=1$, $j=0$,
and $a=0.4999$.
From the figure, we can confirm that $V_{+}$ as a function of $r$
have two extremes for $0\le\mu<1.321$. From Figure
\ref{fig:facpot1}, we can indeed see that $V_+$, in which $\mu=0.7$,
has a local minimum near $r^{-1}=0.15$ and that there is a potential
bound state if we take $\omega\sim0.6878$, which is a correct
eigenfrequency of the bound states. Another important thing is that
$\omega\sim0.6878$ is below $m\Omega$, which can be easily seen from
the factorized potential diagrams. This means that this bound state
is an unstable mode because it satisfies the condition of the
superradiant instability.

\begin{figure}
\centerline{\mbox{\psfig{figure=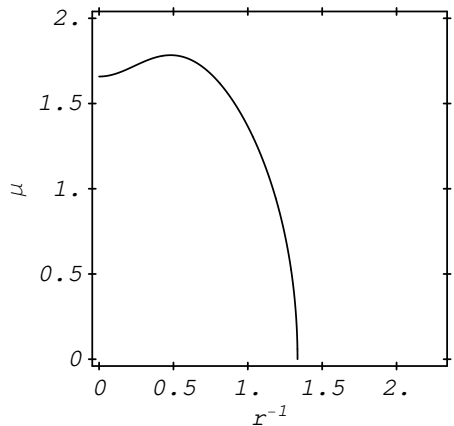,angle=0,width=6cm}}}
\caption{The curve of $dV_+/dr(r^{-1},\mu)=0$ for the parameter
$n=1$, $l=m=1$, $j=0$, and $a=0.9$.\label{fig:facpotmais2}}
\end{figure}

\begin{figure}
\centerline{\mbox{\psfig{figure=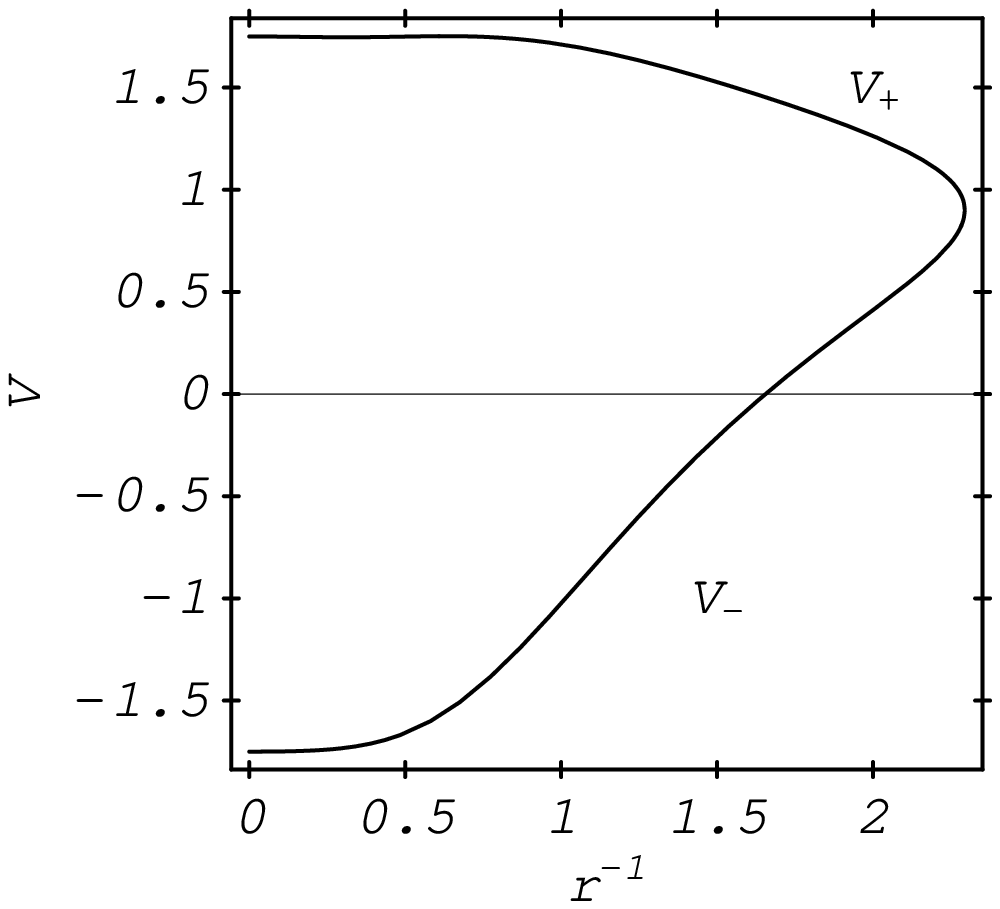,angle=0,width=6cm}
\psfig{figure=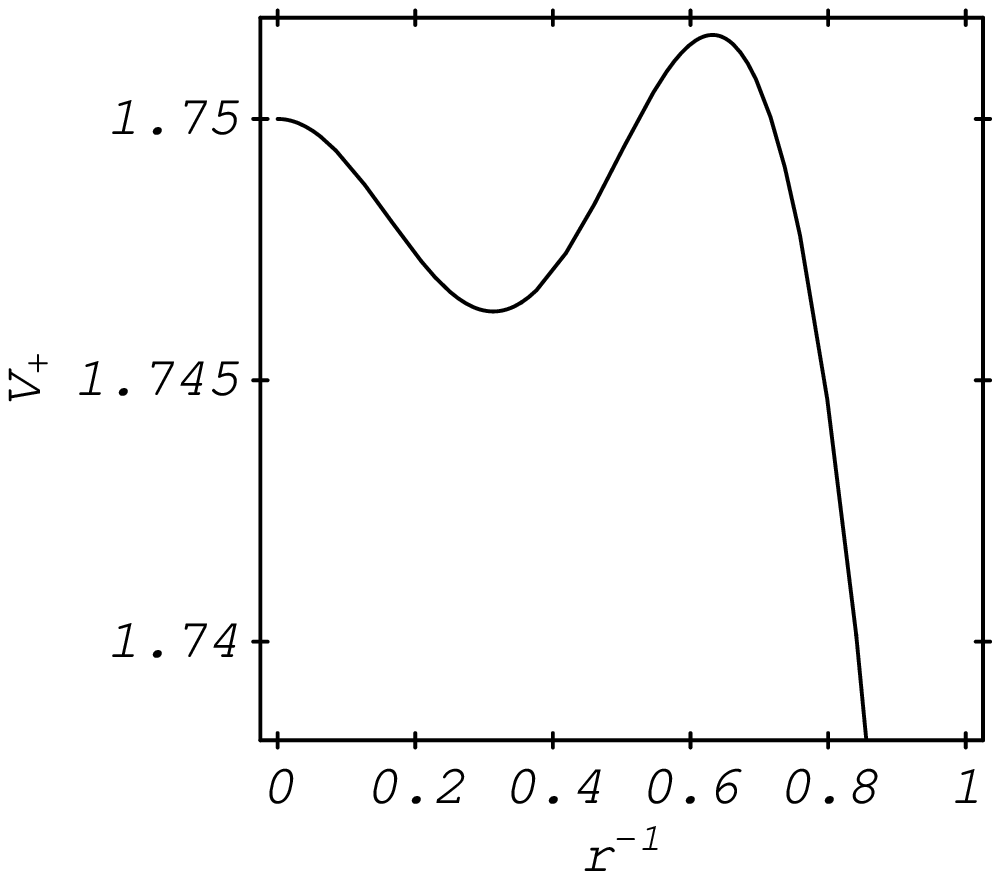,angle=0,width=6cm}}} \caption{The
factorized potentials $V_\pm$ as a function of $r^{-1}$. The
potentials are characterized by $n=1$, $l=m=1$, $j=0$, $\mu=1.75$,
and $a=0.9$ (left panel). The right panel is the same as the left
one, but, a close-up of the potential $V_+$ near
extremes.\label{fig:facpot2}}
\end{figure}

Now, let us move on to higher dimensions, taking $n=1$ to be
specific. Consider the case of $l=m=1$ and $j=0$ as an example. We
are interested in finding out whether or not there can be bound
states in the five dimensional case. In order to see the
distribution of extremes of $V_+$, which give us a criterion for
nonexistence of the bound states, a curve of $dV_{+}/dr(r,\mu)=0$ on
the $r^{-1}$ -- $\mu$ plan is sketched in Figure
\ref{fig:facpotmais2}, where the black hole characterized by $a=0.9$
is taken. In the figure, it is found that if the mass $\mu$
satisfies $1.658\le\mu<1.784$, $dV_{+}/dr$ as a function of $r$ has
two zeros; one corresponds to a local maximum of $V_+$ and the other
a local minimum of $V_+$. It should be emphasized that contrary to
the four dimensional case, there is a critical mass below which no
potential bound state exists. Therefore, there is the possibility
that a bound state exists in the five dimensional case. A
distribution of $V_{+}$ for the case of $\mu=1.75$ is shown in
Figure \ref{fig:facpot2}. We can conclude from the figure that all
the frequencies for potential bound states are always above
$m\Omega$. Therefore, those potential bound states must not be
unstable against the superradiance instability even if they exist.
For other sets of the parameters, we have found similar behavior of
the factorized potential. This means that in the five dimensional
case, there might be bound states but all the potential bound stats
are stable. This analysis is consistent with our numerical studies,
in which we have not found any unstable modes.

For general $n>5$ the trend is the same, there are no unstable bound
states.


\begin{thebibliography}{99}
\bibitem{orbits} Tangherlini \cite{tangher}, who discovered the
higher dimensional Schwarzschild solution, also proved there are no
stable circular geodesics in these spacetimes for ${\rm d}>4$. There
is, to our knowledge, no proof of this statement for higher
dimensional Kerr black holes, but it was shown in \cite{frolov2}
that the five-dimensional Kerr solution doesn't have stable circular
geodesics, along the equatorial plane. It seems very likely that
stable orbits do not exist in the general case.
\end{thebibliography}
\end{document}